\documentclass[twocolumn,showpacs,preprintnumbers,amsmath,amssymb,floatfix]{revtex4}

\usepackage{graphicx}
\usepackage{dcolumn}
\usepackage{bm}

\newcommand{\beq}{\begin{equation}}
\newcommand{\eeq}{\end{equation}}
\newcommand{\bey}{\begin{eqnarray}}
\newcommand{\eey}{\end{eqnarray}}

\begin{document}

\title {Singularity free stars in (2+1) dimensions}

\author{Farook Rahaman}
 \email{rahaman@iucaa.ernet.in}
\affiliation {Department of Mathematics, Jadavpur University',
 Kolkata-700032, India}
 \author{  Ayan Banerjee}
 \email{ ayan_7575@yahoo.co.in}
\affiliation {Department of Mathematics, Jadavpur University',
 Kolkata-700032, India}
\author{Irina Radinschi}
\email{radinschi@yahoo.com} \affiliation {Department of Physics,
"Gh. Asachi" Technical University, Iasi 700050, Romania
 }
 \author{ Sumita Banerjee}
\email{banerjee.sumita.jumath@gmail.com} \affiliation{Dept. of Mathematics,
Adamas Institute of Technology, Barasat, North 24 Parganas -
700126, India}
\author{Soumendranath Ruz}
 \email{ruzfromju@gmail.com}
\affiliation {Department of Physics, Satitara High School, Kandi,
Murshidabad, India - 742170}

\date{\today}

\begin{abstract}

We present some  new types of  non-singular model for anisotropic
stars with constant $\Lambda $ and variable $\Lambda$ based on the
Krori and Barua (KB) metric in $(2+1)$ dimensions. The solutions
obtained here satisfy all the regularity conditions and its simple
analytical form helps us to study the various physical properties
of the configuration.
\end{abstract}


 \maketitle

\section{Introduction}

 The study of (2+1)  dimensional gravity  has become a subject of considerable
interest.  Newtonian theory can not be obtained as a limit of
Einstein's
 theory  in (2+1) dimensional
spacetime. In this case
 gravity is localized that means there is no propagation  of gravity outside the sources. Also,   it is argued that (2+1) dimensional gravity  provides
some new features  towards a better understanding of the
physically relevant (3+1)~dimensional gravity [1-4]. Most of the
studies on this respect are  black hole spacetimes, star    or
cosmological models [5]. However, in recent years various studies
have been done. Rahaman et al[6] have proposed a new model of a
gravastar in (2+1) anti-de Sitter space-time. They have  also
generalized their earlier work on gravastar in (2 + 1) dimensional
anti-de Sitter space-time to (2 + 1) dimensional solution of
charged gravastar [7]. Some authors [8] have also discussed
wormhole solutions  in (2+1) dimensional  spacetime.

In recent past,  Krori and Barua (KB)[9] constructed static,
spherically symmetric solutions  based on a particular choice of
the metric components $g_{00}$ and $g_{11}$ in curvature
coordinates.  Recently, KB's approach was adopted by various
authors for constructing star models [10-15]. These studies are
confined within (3+1) dimensional spacetimes. Therefore, it will
be interesting to search whether nonsingular solutions of KB type
will be existed in (2+1) dimensional fluid sphere. We are looking
forward to get some extra features as our (2+1) dimensional models
include an additional parameter, cosmological constant. In 1999,
Lubo et al [2] has discussed regular (2 + 1) spherically symmetric
solutions, however, there approach were different. Anisotropy was
used in the compact star configuration to allow some interesting
studies and we mention some papers that yield meaningful results
[16]-[19].

In this investigation,   we explore the possibility of applying
the Krori-Barua [9]    metric to describe the interior spacetime
of a
  star in (2+1) dimension.
Nowadays, it is known that the dark energy represents $73\%$ of
the whole mass-energy of our Universe. This conclusion is given by
the Wilkinson Microwave Anisotropy Probe (WMAP) that also
indicates that the dark-energy is causing a speeding of the
expansion of the rate of the universe. The subject of compact
stars is of actuality and under study in the last decades. One of
the  possibility for the formation of  compact anisotropic stars
is to use the cosmological constant. For this reason, we have
considered cosmological constant in our model. We have discussed
two models, one with      constant $\Lambda $ and the other with
 variable $\Lambda $.

The structure of our work is as follows: In section II,  the
non-singular model for anisotropic stars with constant $\Lambda $
based on the Krori and Barua (KB) metric in $(2+1)$ dimensions is
developed. In section III, we have discussed some physical
features of the model. In section IV,  we have  presented the
model with variable $\Lambda $. In Section V,  we have analyzed
some  physical properties of the model given in section IV.
Finally, in section VI,  we have made a conclusion about our work.

\section{Non-singular model for anisotropic   stars with constant
$\Lambda$}

Let us assume that the interior space-time of a   star is
described by the KB   metric

\begin{equation}
ds^2=-e^{2\nu(r)}dt^2 + e^{2\mu(r)}dr^2 +r^2 d\theta^2 ,
\label{eq1}
\end{equation}
with $2\mu(r)=Ar^2$ and $2\nu(r) = Br^2 + C$ where $A$, $B$ and
$C$ are
 arbitrary constants  which will be determined  on the ground of
various physical requirements.
 The energy momentum tensor in the  interior of the anisotropic  star  is assumed in the following standard
 form
\[T_{ij}=diag(\rho,-p_r,-p_t),\]
where $\rho$, $p_r$ and $p_t$ correspond to the energy density,
normal pressure and transverse pressure respectively.

 Therefore, the Einstein field equations
for the metric (1) with constant $\Lambda$ can be written as
(assuming  natural units $G = c=1$)
\begin{eqnarray}
\label{eq2}
 2\pi\rho+ \Lambda &=&
\frac{\mu^{\prime}e^{-2\mu}}{r},
\\
\label{eq3} 2\pi p_r - \Lambda &=& \frac{\nu^{\prime}e^{-2\mu}}{r}
,\\
\label{eq4} 2\pi\ p_t - \Lambda &=&
e^{-2\mu}\left(\nu^{\prime\prime}+\nu^{\prime 2}-
{\nu^\prime\mu^\prime}\right).
\end{eqnarray}

Now, plugging  the metric (1) in equations (2) - (4), we get the
following expressions of energy density $\rho$, normal pressure
$p_{r}$, tangential pressure $p_{t}$   as
 \begin{eqnarray}
\rho &=& \frac{1}{2\pi}\left[Ae^{-Ar^2}-\Lambda\right] , \label{eq5}\\
p_{r} &=& \frac{1}{2\pi}\left[Be^{-Ar^2}+\Lambda\right] , \label{eq6}\\
p_{t} &=&
\frac{1}{2\pi}\left[e^{-Ar^2}\left(B^2r^2+B-ABr^2\right)+\Lambda\right]
\label{eq7}.
\end{eqnarray}

 Using equations (5) - (7),  the equations of state
(EOS) parameters corresponding to radial and transverse directions
are written as
\begin{equation}
\label{eq11} \omega_r(r)   =
\frac{Be^{-Ar^2}+\Lambda}{Ae^{-Ar^2}-\Lambda}
\end{equation}
\begin{equation}
\label{eq12} \omega_t(r) =
\frac{e^{-Ar^2}\left(B^2r^2+B-ABr^2\right)+\Lambda}{Ae^{-Ar^2}-\Lambda}.
\end{equation}

\section{PHYSICAL ANALYSIS}

In this section we will discuss the following features of our
model:

\subsection{Matching Conditions}

The  exterior ($p = \rho = 0$) solution corresponds to a static,
  BTZ black hole is written in the following form as
\begin{equation}
ds^{2}=-\left(-M_0 - \Lambda{r^2} \right)dt^2 + \left(-M_0 -
\Lambda{r^2} \right)^{-1}dr^2 + r^2d\theta^2, \label{eq20}
\end{equation}
Here we match our interior metric to the exterior BTZ  metric and
as a consequence, we get
\begin{equation}
A = -\frac{1}{R^2}{\ln(K^2R^2- M_{0}) },
\end{equation}
\begin{equation}
B=\frac{K^2}{K^2R^2-M_0},
\end{equation}
\begin{equation}
C = \frac{K^2}{K^2R^2-M_0}-\ln(K^2R^2- M_{0}).
\end{equation}

\subsection{Regularity at the centre}

In this analysis, we consider anti-de Sitter space-time and to
ensure that cosmological constant is  always negative, we write
$\Lambda=-K{^2}$. Since the radial EOS  is always less than unity,
therefore equation (8) at once indicates that the cosmological
constant should be negative.

Now, we find the central density and  central pressures ( radial
and transverse ) as
\begin{equation}
\rho{_0}=\rho (r=0)=\frac{1}{2\pi}\left(A+K^2\right)
\end{equation}
\begin{equation}
p_r(r=0) = p_t(r=0) = p{_0}=\frac{1}{2\pi}\left(B-K^2\right)
\end{equation}
From the equations (5) and (6) , we find
\[ \frac{d\rho}{dr}= -\frac{A^2}{\pi}re^{-Ar^2} < 0,\]
~~~and~~~
\[ \frac{d p_r}{dr}= -\frac{AB}{\pi}re^{-Ar^2} <0.\]
which gives a meaningful result that density and pressure are
decreasing function of r.
 The above equations imply
at $r=0$,
\begin{equation} \frac{d\rho}{dr}=0,~~~~ \frac{dp_r}{dr} = 0,\end{equation}
\[ \frac{d^2 \rho}{dr^2}=-\frac{A^2}{\pi} < 0, \]
~~~and~~~
\[\frac{d^2 p_r}{dr^2} = -\frac{AB}{\pi}< 0.\]
which support maximality of central density and radial central
pressure.

For our model, the measure of anisotropy, $\Delta  = p_{t}-p_{r}$,
  is given by
\begin{equation}
\label{eq13} \Delta  =
 \frac{B(B-A)}{2\pi}r^2e^{-Ar^2}.
\end{equation}
 The anisotropy, as expected, vanishes at
the centre.

 The energy density and the
two pressures are continuous function of radial coordinate r that
means they are well behaved in the interior of the stellar
configuration.  The radius of star is obtained by letting
$p_r(r=R)=0$ , which gives
\begin{equation}
R=\sqrt{\frac{1}{A} \ln\left(\frac{B}{K^2}\right)}.
\end{equation}

 We have considered the data from a 3 spatial dimensional
object in order to fix the constants of the model. Although it is
not very clear the physical meaning by considering stelar objects
of 2 spatial dimensions, however, if an observer is sitting in the
plane, $\theta~ = ~constant$ , then he sees all characteristics as
a (2+1) dimensional picture. So apparently, one can use all the
  data which are more or less the same for both spacetimes.

 Therefore we have used the data from  X ray buster 4U
1820-30 to calculate the corresponding constants.  It is known
that  the mass of X ray buster 4U 1820-30 is  2.25 $M_{\odot}  $
and radius 10 km.  To understand  the physical  behavior of the
solutions of our model, we have assumed certain values of
$\Lambda$. As we considered the anti de-Sitter spacetime, we have
used negative values, say, $\Lambda =-.035, -.038,  -0.04$ ( see
reference [20] ).

 Using the data from X ray buster 4U 1820-30,  we have
obtained the values of the constants A and B via equations (11)
and (12)  for different values of $\Lambda$ in units of $km^{-2}$
( see table 1).

\begin{table*}
\centering
\begin{minipage}{140mm}
\caption{Values of the constants A, B  for   X ray buster 4U
1820-30 for different values of $\Lambda$.}\label{tbl-1}
\begin{tabular}{@{}lrrrr@{}}
\hline $\Lambda$   &   {$M$ } & ~~~~~~~~~~~~{$R$ (km)} ~~& {A} & {B}\\
\hline
-0.035 &  ~~~~~~ 2.25 $M_{\odot}  $  &10 & ~~~0.017078 &~~~~~ 0.19310\\
 -0.038 & ~~~~~~  2.25 $M_{\odot}  $  &10 & ~~~0.007313 & ~~~~~0.07896\\
-0.04 & ~~~~~~  2.25 $M_{\odot}  $  &10 & ~~~0.003834 & ~~~~~0.05869\\
\hline
\end{tabular}
\end{minipage}
\end{table*}

Plugging in G and c in the relevant equations, we have calculated
the central density $\rho_0 = 9.49 \times 10^{15} gm~ cm^{-3}$,
surface density $\rho_R =   9.23 \times 10^{15} gm~ cm^{-3}$,
central pressure $p_r(r = 0) = p_t(r = 0) = 3.28 \times 10^{36}
~dyne ~cm^{-2}$ \textbf{for $\Lambda = -.04$.}

 We draw the figures for the density variation and
pressures at the stellar interior of X ray buster 4U 1820-30 ( see
figures 1  and 2  ). The figure 3 indicates that $\Delta
> 0 $  i.e. the `anisotropy' is
directed outward. Thus  our model exerts a repulsive `anisotropic'
force ($\Delta > 0$) which  allows the construction of more
massive distributions. In figure 2, one can note that transverse
pressure is increasing towards the surface. These types of stars
are  composed of unknown matters whose density is above the normal
nuclear density, $\rho_n \sim 4.6\times10^{14} gm~cm^{-3}$. As a
result, these highly compact stars are anisotropic in nature and
peculiar phenomena may be happened there.

\begin{figure}
\centering
\includegraphics[scale=.4]{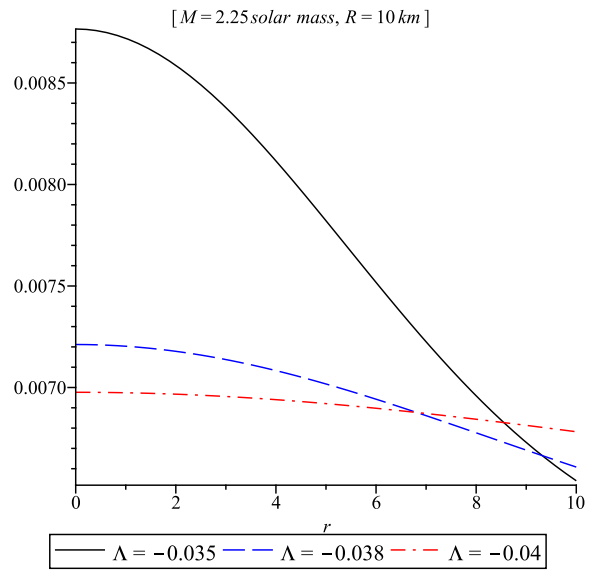}
\caption{Density ( along vertical axis ) variation  at the stellar
interior of X ray buster 4U 1820-30 of mass 2.25 $M_{\odot}  $ and
radius 10 km for different values of  $\Lambda$.} \label{fig:1}
\end{figure}
\begin{figure}
\centering
\includegraphics[scale=.4]{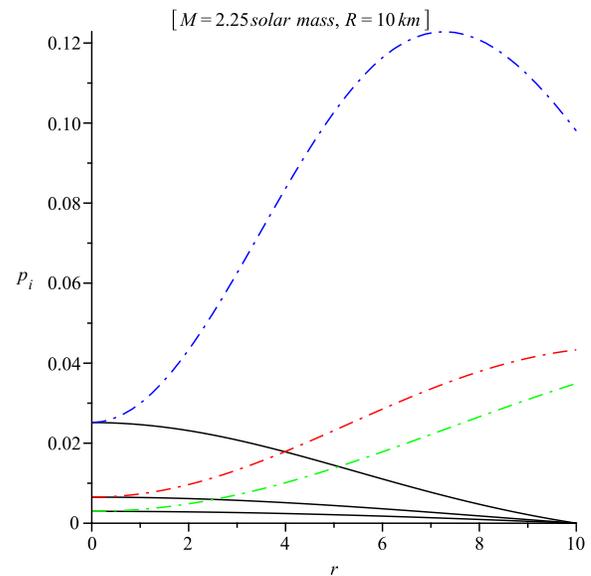}
\caption{Radial and transverse pressures variation at the stellar
interior X ray buster 4U 1820-30 of mass 2.25 $M_{\odot}   $ and
radius 10 km for different values of  $\Lambda$. Solid lines and
chain lines  indicate radial  and transverse pressures
respectively. }
    \label{fig:2}
\end{figure}
\begin{figure}
\centering
\includegraphics[scale=.4]{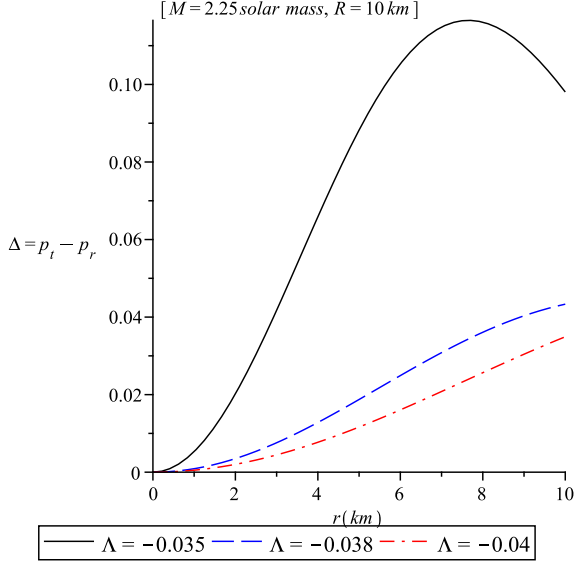}
\caption{The anisotropic behaviour   at the stellar interior X ray
buster 4U 1820-30 of mass 2.25 $M_{\odot} $ and radius 10 km for
different values of  $\Lambda$.}
    \label{fig:2}
\end{figure}

\subsection{TOV Equation}
 The generalized TOV
equation for an anisotropic fluid distribution is given by
\begin{equation}
\label{eq24} \frac{d}{dr}\left(p_r -\frac{\Lambda}{2\pi} \right) +
\nu^\prime\left(\rho +p_r\right) + \frac{1}{r}\left(p_r -
p_t\right) = 0.
\end{equation}

According to  Ponce de Le\'{o}n [21] suggestion, one can rewrite
 the above TOV equation as
\begin{eqnarray}
-\frac{M_G\left(\rho+p_r\right)}{r^2}e^{\frac{\mu-\nu}{2}}-\frac{d}{dr}
\left(p_r -\frac{\Lambda}{2\pi} \right)
+\frac{1}{r}\left(p_t-p_r\right)=0, \label{eq25}
\end{eqnarray}
where $M_G=M_G(r)$ is the gravitational mass inside a sphere of
radius $r$ and is given by
\begin{equation}
M_G(r)=r^2e^{\frac{\nu-\mu}{2}}\nu^{\prime},\label{eq26}
\end{equation}
[ This can be derived from the Tolman-Whittaker formula using
Einstein field equations.]

This modified form of  TOV equation delineates  the equilibrium
condition for the X ray buster 4U 1820-30 subject to the
gravitational and hydrostatic plus another force due to the
anisotropic nature of the stellar object. Now, we write the above
equation  as
\begin{equation}
 F_g+ F_h + F_a=0,\label{eq27}
\end{equation}
where

\begin{eqnarray}
F_g &=& -B r\left(\rho+p_r\right), \label{eq28}\\
F_h &=& -\frac{d}{dr}\left(p_r -\frac{\Lambda}{2\pi} \right) \nonumber\\
&=& \frac{AB}{\pi}re^{-Ar^2},
 \label{eq29}\\
F_a &=& \frac{1}{r}\left(p_t -p_r\right). \label{eq30}
\end{eqnarray}

  The profiles of $F_g$, $F_h$ and $F_a$ for our
chosen source are shown in Fig. 4. The figure provides the
information  about the static equilibrium   due to the combined
effect of pressure anisotropy, gravitational and hydrostatic
forces.

\begin{figure}
\centering
\includegraphics[scale=.35]{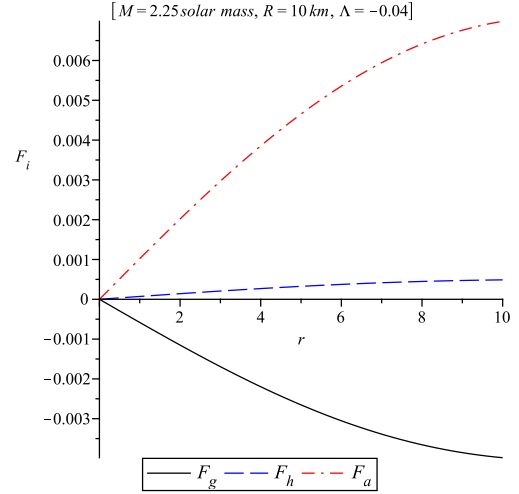}
\caption{Three different forces acting on fluid elements of X ray
buster 4U 1820-30 of mass 2.25 $M_{\odot} $ and radius 10 km in
static equilibrium is shown against $r$ with $\Lambda = -.04$. }
    \label{fig:2}
\end{figure}

\subsection{Maximum mass-radius relation }

We calculate the mass $m(r)$   within a  radial distance $r$
  as
\begin{equation}
m(r) = \int_0^r 2\pi \rho \tilde{r}d\tilde{r} = \frac{1}{2} [1+K^2
r^2- e^{-Ar^2}]  .\label{eq8}
\end{equation}
In Fig.~5, we have plotted this mass to radius relation. One can
note that the upper bound on the mass in our model can be written
as
\begin{equation}
2m(r) \equiv 1+K^2 r^2- e^{-Ar^2} \leq 1+K^2r^2
-e^{-AR^2},\label{eq31}
\end{equation}
which implies
\begin{equation}
\left(\frac{m(r)}{r}\right)_{max} \equiv \frac{M}{R}  \leq
\frac{1+ K^2 R^2 -e^{-AR^2}}{2R}.\label{eq32}
\end{equation}
  The
plot  $\frac{m(r)}{r}$ against $r$ (see Fig.6) indicates that the
ratio $\frac{m(r)}{r}$ is an increasing function of the radial
parameter. It is interesting to note  that the constraint on the
maximum allowed mass-radius ratio in our case falls within the
limit   to the (3+1) dimensional case of  isotropic fluid sphere
i.e., $\left(\frac{m(r)}{r}\right)_{max} = 0.216 < \frac{4}{9}$ (
here we have used $\Lambda = -.04$ ).
\begin{figure}
\vspace{0.2cm} \includegraphics[width=0.4\textwidth]{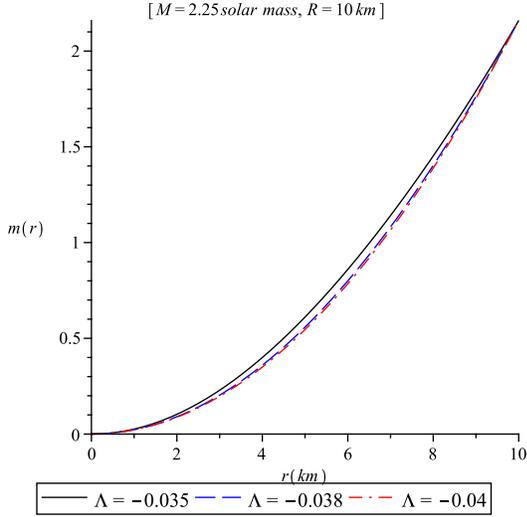}
\caption{Variation of mass function at stellar interior for
different values of  $\Lambda$.} \label{fig:3}
\end{figure}
\begin{figure}
\vspace{0.2cm} \includegraphics[width=0.45\textwidth]{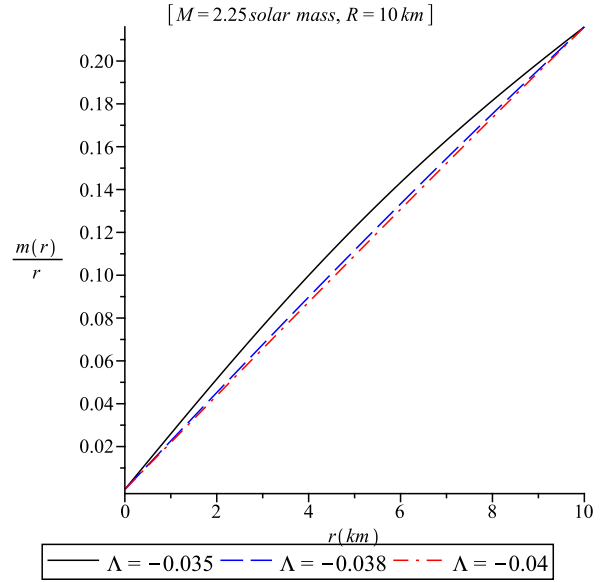}
\caption{  Variation of $\frac{m(r)}{r}$  is shown against r for
different values of  $\Lambda$.  } \label{fig:3}
\end{figure}

\subsection{Compactness and redshift}
From the above mass function (26), we obtain the compactness of
the star as
\begin{equation}
\label{eq33} u= \frac{ m(r)} {r}=  \frac{1}{2r}\left( 1+K^2 r^2-
e^{-Ar^2}
 \right),
\end{equation}
and  correspondingly the surface redshift ($Z_s$)  is given by
\begin{equation}
\label{eq34} Z_s= ( 1-2 u)^{-\frac{1}{2}} - 1,
\end{equation}
where
\begin{equation}
\label{eq35} Z_s=  \left[  1-  \frac{1}{r}\left( 1+K^2 r^2-
e^{-Ar^2}
 \right)\right]^{-\frac{1}{2}} - 1 .
\end{equation}
Thus, the maximum surface redshift of our (2+1) dimensional  star
of radius 10 km can be found as  $Z_s =  0.328$  ( here we have
used $\Lambda = -.04$ ).\\
The figure 7 indicates the variation of redshift function $Z_s$
against r for different values of  $ \Lambda $.
\begin{figure}
\begin{center}
\vspace{0.2cm} \includegraphics[width=0.45\textwidth]{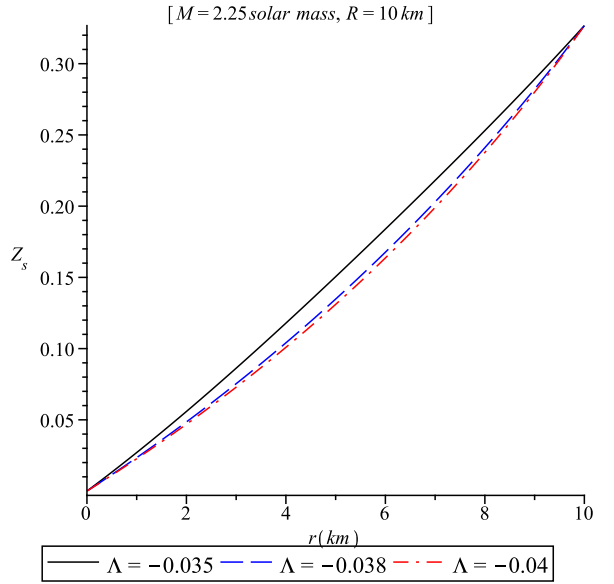}
\end{center}
\caption{The variation of redshift function $Z_s$ is shown against
r for different values of  $\Lambda$.} \label{fig:4}
\end{figure}
\\
\\
\\
\\
\\

\section{Variable $\Lambda $}

Now, we consider the model taking the cosmological constant as
radial dependence i.e. $\Lambda ~=~ \Lambda(r) ~=~\Lambda_r$
(say).

To get the physically acceptable stellar models, we assume that
the radial pressure of the compact star is proportional to the
matter density i.e.
\begin{equation}
\label{eq5} p_r = m\rho,~~~ m>0,
\end{equation}
 where $m$ is the equation of state parameter.

Now, using the above equation of state and equations (1) - (4), we
get the energy density $\rho$, normal pressure $p_{r}$, tangential
pressure $p_{t}$ and cosmological parameter $\Lambda_{r}$,
respectively as
 \begin{eqnarray}
\rho &=& \frac{(A+B)}{2\pi (m+1)}e^{-Ar^2} >0, \label{eq6}\\
p_{r} &=& \frac{m(A+B)}{2\pi (m+1)}e^{-Ar^2} >0, \label{eq7}\\
p_{t} &=& \frac{e^{-Ar^2}}{2\pi}[Br^2(B-A)+\frac{m(A+B)}{1+m}].
 \label{eq8}\\
\Lambda_{r} &=& \frac{e^{-Ar^2}}{m+1}\left[mA-B\right].
\end{eqnarray}

Also, the equation of state (EOS) parameters corresponding to
normal and transverse directions can be written as
\begin{equation}
\label{eq11} \omega_r(r)   =  m,
\end{equation}
\begin{equation}
\label{eq12} \omega_t(r) = m+\frac{(m+1)(B-A)Br^2}{(A+B)}.
\end{equation}

\section{PHYSICAL ANALYSIS}

In this section we will discuss the following features of our
model:

One can see from equations (33) and (34) that
\[ \frac{d\rho}{dr}= -\left[\frac{A(A+B)}{\pi (m+1)}re^{-Ar^2}\right] < 0,\]
~~~and~~~
\[ \frac{d p_r}{dr}= -\left[\frac{m(A+B)A}{\pi (m+1)}re^{-Ar^2}\right] < 0.\]
Also,  at $r=0$, our model provides
\[ \frac{d\rho}{dr}=0,~~ \frac{dp_r}{dr} = 0, \]
\[ \frac{d^2 \rho}{dr^2}=-\frac{A(A+B)}{\pi (m+1)}  < 0, \]
~~~and~~~
\[\frac{d^2 p_r}{dr^2} = -\frac{mA(A+B)}{\pi (m+1)} < 0.\]
which indicate maximality of central density and central
pressure.\\ The central density and radial central pressures  are
given by
\begin{equation}
\rho{_0}=\rho (r=0)=\frac{\left(A+B\right)}{2\pi(m+1)}
\end{equation}
\begin{equation}
p_r(r=0) = p_t(r=0) = p{_0}=\frac{m(A+B)}{2 \pi(m+1)}
\end{equation}

 It is obvious that the radial pressure vanishes at the surface
 i.e. at r=R, $p_r ( r=R) =0$ where R is the radius of the star.
 Equation (34) implies either m = 0 or A = - B. But, A = - B implies
 $\rho =0$, which is not possible, therefore, we should take m=0. In other words, in (2+1) dimensional KB spacetime with variable
$\Lambda$, only dust anisotropic model exists.

The measure of anisotropy, $\Delta  = p_{t}-p_{r}$, in this model
is obtained as
\begin{equation}
\label{eq13} \Delta  =
 \frac{B(B-A)}{2\pi}r^2e^{-Ar^2}.
\end{equation}
At the centre $\Delta = 0$ i.e. anisotropy dies out as expected. \\

 We match the interior metric to the exterior BTZ matric
and get fortunately the same values of  constants A, B and C given
in equations (11)-(13).

As before, we have considered X ray buster 4U 1820-30  and have
chosen the values of the parameters A and B as A=0.003834 and B =
0.05869 to obtain
 the central density and surface density which are given by $\rho_0 = 1.36 \times 10^{16} gm~
cm^{-3}$  and $\rho_R =   9.234 \times 10^{15} gm~ cm^{-3}$
respectively.
\\
In this case the TOV equation is written as
\begin{equation}
 F_g+ F_h + F_a=0,\label{eq27}
\end{equation}
where

\begin{eqnarray}
F_g &=& -B r\left(\rho \right), \nonumber\\
F_h &=& -\frac{d}{dr}\left(  -\frac{\Lambda_r}{2\pi} \right) \nonumber\\
 \label{eq29}
F_a &=& \frac{1}{r}\left(p_t  \right). \label{eq30}
\end{eqnarray}

We note that here the variable $\Lambda$ contributes to the
hydrostatic force and transverse pressure provide the effect of
pressure anisotropy.

 The profiles of $F_g$, $F_h$ and $F_a$ for our
chosen source are shown in Fig. 8. The figure provides the
information  of the static equilibrium   due to the combined
effect of pressure anisotropy, gravitational and hydrostatic
forces.
\begin{figure}
\centering
\includegraphics[scale=.35]{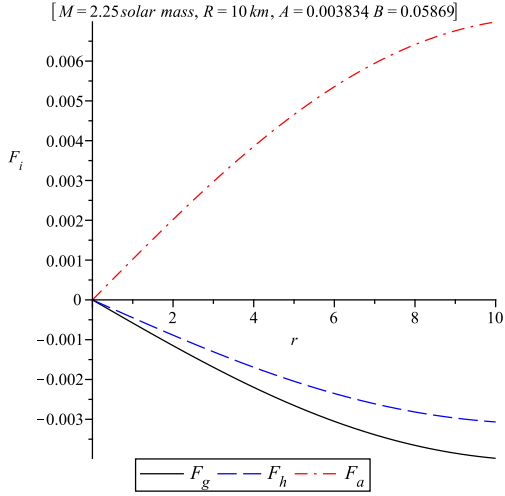}
\caption{Three different forces acting on the fluid elements of X
ray buster 4U 1820-30 of mass 2.25 $M_{\odot} $ and radius 10 km
in static equilibrium is shown against $r$. }
    \label{fig:2}
\end{figure}

The effective mass $m(r)$   within a  radial distance $r$   is
defined
  as
 \[
m_{eff}(r) = \int_0^r 2\pi \left[ \rho + \frac{\Lambda_r }{2
\pi}\right] \tilde{r}d\tilde{r}\] Hence, we get
\begin{equation}
m_{eff}(r) = \frac{1}{2} [1 - e^{-Ar^2}]  .\label{eq8}
\end{equation}
In Fig.~9, we plot this mass to radius relation. One can note that
the upper bound on the mass in our model can be written as
\begin{equation}
2m_{eff}(r) \equiv 1- e^{-Ar^2} \leq 1 -e^{-AR^2},\label{eq31}
\end{equation}
which implies
\begin{equation}
\left(\frac{m_{eff}(r)}{r}\right)_{max} \equiv \frac{M_{eff}}{R}
\leq \frac{1 -e^{-AR^2}}{2R}.\label{eq32}
\end{equation}
  The
plot  $\frac{m_{eff}(r)}{r}$ against $r$ (see Fig.10) indicates
that the ratio $\frac{m_{eff}(r)}{r}$ is an increasing function of
the radial parameter. Again, we see that maximum allowed
mass-radius ratio in our case falls within the limit   to the
(3+1) dimensional case of isotropic fluid sphere i.e.,
$\left(\frac{m_{eff}(r)}{r}\right)_{max} = 0.01598 < \frac{4}{9}$
( here we have used A=0.003834 and B = 0.05869 ).
\begin{figure}
\vspace{0.2cm} \includegraphics[width=0.45\textwidth]{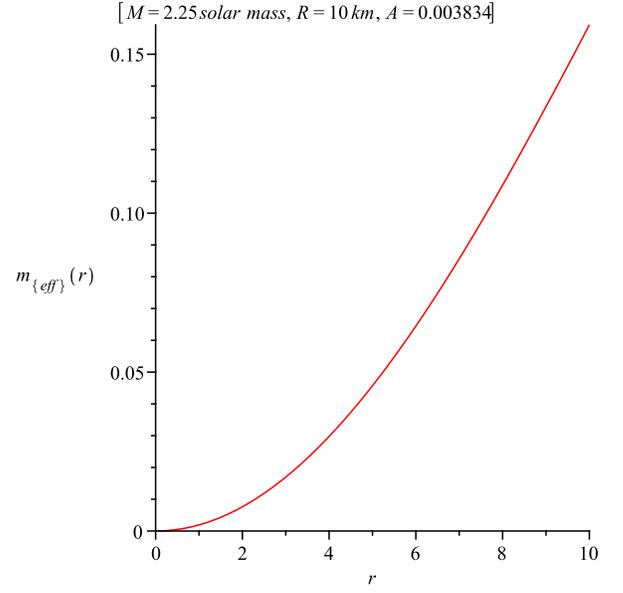}
\caption{Variation of mass function against r  at stellar
interior.} \label{fig:3}
\end{figure}
\begin{figure}
\vspace{0.2cm} \includegraphics[width=0.45\textwidth]{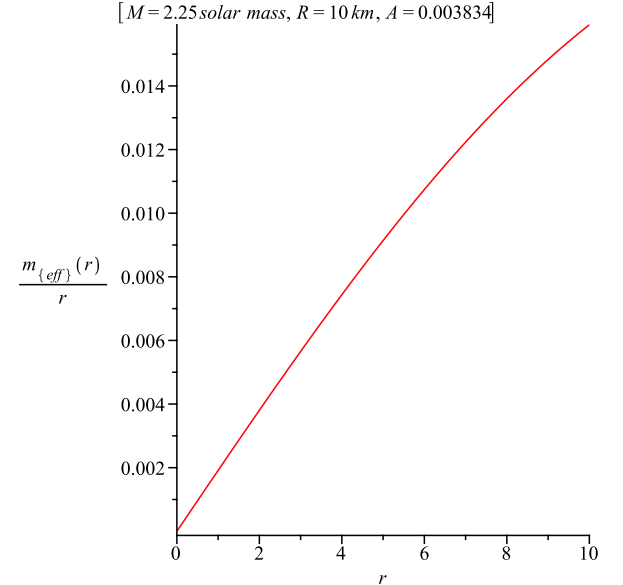}
\caption{Variation   of $\frac{m _{eff}}{r}$  is shown against r
at stellar interior. } \label{fig:3}
\end{figure}

The compactness of the star is given as
\begin{equation}
\label{eq33} u= \frac{ m(r)} {r}=  \frac{1}{2r}\left( 1 -
e^{-Ar^2}
 \right),
\end{equation}
and  correspondingly the surface redshift ($Z_s$)  is given by
\begin{equation}
\label{eq34} Z_s= ( 1-2 u)^{-\frac{1}{2}} - 1,
\end{equation}
where
\begin{equation}
\label{eq35} Z_s=  \left[  1-  \frac{1}{r}\left( 1- e^{-Ar^2}
 \right)\right]^{-\frac{1}{2}} - 1 .
\end{equation}
Thus, the maximum surface redshift of our (2+1) dimensional  star
of radius 10 km can be found as  $Z_s =  0.0164$ ( here we have
used A=0.003834 and B = 0.05869  ). The figure 11 indicates the
variation of redshift function $Z_s$ against r.
\begin{figure}
\begin{center}
\vspace{0.2cm} \includegraphics[width=0.45\textwidth]{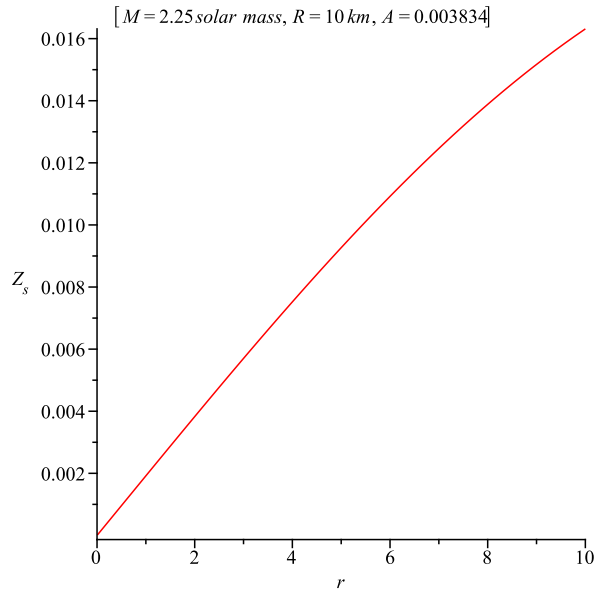}
\end{center}
\caption{The variation of redshift function $Z_s$ is shown against
r at stellar interior. } \label{fig:4}
\end{figure}

\section{Concluding Remarks}

By taking the Krori and Barua  metric  as input and treating the
matter content  as anisotropic in nature , we have obtained some
new types non-singular solutions for   stars with constant
$\Lambda $ and variable $\Lambda$. In this phenomenological model,
we have employed physical data of the X ray buster 4U 1820-30 in
our models. It is found that the central density $\rho_0 = 9.49
\times 10^{15} gm~ cm^{-3}$ and  surface density $\rho_R =   9.23
\times 10^{15} gm~ cm^{-3}$ for the first case which are beyond
  the normal nuclear density. For the later case, densities are
more than the former one. The models are attainable  in  static
equilibrium  conditions  due to the combined effect of pressure
anisotropy, gravitational and hydrostatic forces. One can note
that the maximum allowed mass-radius ratio in our cases falls
within the limit   to the (3+1) dimensional case of isotropic
fluid sphere i.e., $\left(\frac{m_{eff}(r)}{r}\right)_{max}   <
\frac{4}{9}$. We have seen that  in (2+1) dimensional KB spacetime
with variable $\Lambda$, only dust anisotropic model exists. This
  feature is absent in (3+1) dimension.
Investigation on full collapsing model of a $(2+1)$ dimensional
star   is
   beyond the scope of this analysis. We would like to perform
   this study in the future.

\section*{Acknowledgments}

 FR wishes to thank the authorities of the
Inter-University Centre for Astronomy and Astrophysics, Pune,
India for providing the Visiting Associateship under which a part
of this work was carried out. FR is also thankful to PURSE, DST
and UGC, Govt. of India, under post doctoral research award, for
providing financial support.

\end{document}